# Network Generations and the Security Challenge in IoT Applications

Mahmoud S. Fayed

College of Computer and Information Sciences

King Saud University, Riyadh, Saudi Arabia

## Abstract

Networks exist all around on the planet, and inside the brain of every living organism. On the city streets we see transport networks, inside homes and organizations we see water piping networks. And in the digital information age we see the rise of the computer networks and the mobile networks and their upgrade from generation to generation. We need networks for efficient sharing of resources and as human beings the most valuable thing that we can share along the time from age to age is our data that could represent our knowledge and information represented through text, voice, image, and video. In this paper we provide a quick overview on network generation and IoT applications. Also, we point to the security threats in these generations.

## 1. Importance of Networks

Imagine a painter who exist on the earth five thousand years ago. He needs to draw on stones to save his images for the next generations. Transferring these stones from place to place is very hard and time consuming. Spreading it all over the world means drawing it again and again. Continuous cleaning of these paintings to looks nice is required [1-4].

In the current digital age, it is just one click and the data will be accessible everywhere without any physical effort and it could be available for the next generations, infinite years from now. [5] For example, in 2020 GitHub is storing a backup for the open-source projects source code in the Frozen Arctic to be accessible 1000 years from now even if nuclear holocaust happened, the survivor could get it [6].

In the current age, mobile networks and different routing protocols enabled us to get the real power of networking and sharing resources, also this raised the security challenges to very high levels [7-9].

In this paper we review the evolution of the mobile networks from the first generation to the six generation and highlights the security challenges of 6G IoT applications. The remainder of this paper is organized as follows. Section 2 describes the 1G generation; Section 3 describes the 2G generation. Section 4 describes the 3G generation. Section 5 describes the 4G generation. Section 6 describes the 5G generation and the 5G IoT security. Section 7 describes the 6G generation and the 6G IoT security. Section 8 describes Programming Without Coding Technology. Finally concluding remarks are given in Section 9.

## 2. The 1G Generation

Back in 1979 with the start of the 1G generation of commercial Mobile Networks in Japan by Nippon Telegraph and Telephone, the wireless mobile network provided basic voice service using Frequency Division Multiple Access (FDMA) and Analog-based protocols. [10]

In this generation, the network users could only make phone calls without the ability to send text messages. One of the big issues in this generation is related to the security of the conversations. The network architecture suffers from an obvious threat where an attacker could hear and record the conversations because no encryption is used at all. [11]

This generation works at speed from 2.4 Kbps to 14.4 Kbps and does not provide any internet service. This generation uses frequencies around 900 MHz [12]

## 3. The 2G Generation

In 1991 the Radiolinja started the second generation of mobile networks in Finland. In the generation the Radio signals are digitals. The technology uses Time Division Multiple Access (TDMA) and support data services where the users could send and receive SMS text messages. The conversations are encrypted which increase the security in the network architecture. The technology comes with more capacity and support more users. [13]

In 1997 General Packet Radio Service (GPRS) was introduced by the European Telecommunications Standard Institute (ETSI) and users can send the emails. GPRS support data rated from 56 to 114Kbps. The second-generation mobile networks combined with GPRS sometimes called 2.5G to indicate that the system implement packaged-switched domain beside circuit-switched domain. The ETSI developed the Global System for Mobile Communication (GSM) standard for this generation. The Enhanced GPRS (EDGE) provide more speed than GPRS and three times faster that supports 384 Kbps (theoretical speed) by using 8 Phase-Shift Keying (PSK) encoding called 2.75G [14-15]

## 4. The 3G Generation

In 2001, the third generation was introduced that provides at least 144 Kbps of speed and support mobile internet access. Upgrades like 3.5G and 3.75G provide better speed of more Mbps to smartphones.

This generation uses code-division multiple access (CDMA) and provide better security by providing users and uses the KASUMI block cipher. The popular standards for this generation are the Universal Mobile Telecommunication System (UMTS) and the CDMA2000 system. [16]

The introduction of wideband code division multiple access (W-CDMA) protocols allowed UMTS to accelerate the data transfer. The high-speed downlink packet access (HSDPA) and high-speed uplink packet access (HSUPA) extended the performance and improvement. This supports stable video calls on smartphones. [17-19]

During this generation, the smartphones market watched the rise of Android and iOS mobile phones. [20-21]

## 5. The 4G Generation

The 4G generation is commercialized in 2009 that uses the Long-Term Evolution (LTE) developed by 3$^{rd}$ generation partnership project (3GPP) and provide higher transmission of multimedia data like videos and reduce latency. This standard is based on the technologies of GSM/EDGE and UMTS/HPSA. Each country may use a different LTE frequencies and bands requires multi-bands phones.

The 4G networks is not compatible with 2G and 3D and it works on separate radio spectrum. It does not support circuit-switched telephony system. This generation uses internet protocols like voice over IP (VoIP). The evolved standard is LTE Advanced (LTE-A). The 4G is faster than 3G and works at 100 Mbps speed. It has the capacity up to 1 Gbps. This generation uses Orthogonal frequency-division multiple access (OFDMA) digital modulation. 4.5G include technologies as LTE Advanced Pro & MIMO (multiple-input multiple-output) to support higher data rates [22-24].

## 6. The 5th Generation (5G)

The companies started deploying 5G in 2019. This generation provide higher speeds up to 10 Gbps. This increases the possible applications related to Internet of Things (IoT) that needs to connect numerous devices to the internet at the same time to achieve the speed of many Gbps it requires to work at high-frequency bands like 25-39 GHz. The 5G standard started by ITU IMT-2020 then 3GPP selected 5G NR (New Radio) standard and LTE to be submitted for IMT-2020 (International Mobile Telecommunications-2020). In 2020, smartphones like Samsung Galaxy S20, Nokia 8.3 5G and Apple iPhone 12 support connectivity over 5G networks.

Unlike previous generations that focus on the consumer, this generation is designed for the industry revolution that require higher requirements for advanced applications like large-scale IoT applications [25-27].

### 5G-IoT Security

The 5G networks will connect around 7 trillion IoT devices according to an expectation by the 5G-Public Private Partnership (5G-PPP). Security requirements during the usage of these devices and the privacy of the users while providing reliable access everywhere increase the security challenges.

Some of the goals of 5G includes, very high throughput (1-20 Gbps), Ultra low latency (<1ms), 1000x bandwidth per unit area, massive connectivity, High availability, dense coverage, and low energy consumption.

The current technology indicates that the world has gone mobile and there is 10x growth in mobile traffic from 2013 to 2019. Seen the rise of cloud computing, increase in the web traffic driven by internet video, which represent over 57% of the traffic in 2018 and a lot of interest in Internet of Everything (People, Process, Data & Things).

According to ITU, 5G will cover three use-cases, enhanced mobile broadband, ultra-reliable low latency communications (useful for factory automation and robotics too), massive machine type communications.

There are 5G non-standalone solution (NSA) for some initial use cases and 5G standalone (SA) Solution.

Cisco identified five areas to 5G security requirements by stopping threats at the edge, protect users wherever they are, control who gets intro your network, simplify network segmentation, and find and contain problems fast. [28]

Using 5G network, cloud computing, software defined networking (SDN), and network function virtualization (VFV) introduces a lot of security threats. The security threats include DoS attack, Hijacking attacks, signaling storms, resource (slice) theft, configuration attacks, saturation attacks, penetration attacks, User identity theft, TCP level attacks, man-in-the-middle attack, reset and IP spoofing, scanning attacks, security keys exposure, semantic information attacks, timing attacks, boundary attacks and IMSI catching attacks.

The security solutions include Dos and DDoS detection, configuration verification, access control, traffic isolation, link security, identity verification, identity security, location security, IMSI security, Mobile terminal security, Integrity verification, HX-DoS mitigation, and Service access control. New security solutions that use Artificial intelligence and context awareness are required to handle the high traffic produced by IoT devices.

Organizations for standards like 3GPP, 5GPPP, IETF, NGMN and ETSI focus on different major security areas. 3GPP & 5GPP focus on the security architecture and the subscriber privacy. IETF focus on security solutions for massive IoT devices. NGMN focus on network slicing and MEC security. ETSI focus on security architecture and MEC security and privacy. [29-30]

The 5G IoT includes many application domains like Internet of Vehicles (IoV), Smart Healthcare, Smart Home, Unmanned aerial vehicles (UAVs), Smart City and Industry 4.0. due to 5G IoT networks have specific characteristics like Mobility (could cause frequent authentication requests), Massive IoT with resources constraints (A problem in secret key distribution and using advanced cryptography), Heterogenous hierarchical architecture (protecting communication is a problem). Physical layer threats in 5G IoT includes Eavesdropping (Interception & Traffic Analysis), Contaminating (Pilot and Feedback), Spoofing (Identity spoofing and Sybil attacks), Jamming (Pilot, Proactive & Reactive) [31].

5G and massive usage of IoT devices and applications introduces extra challenges related to the security requirements and performance trade-offs.

This opens some research questions like

Is public key cryptography (PKC) mandatory in IoT? And what are the use-cases? And the alternatives.

NIST considered standardization of light weight ciphers only, and 3GPP still too far from decision making [32].

The 3GPP is required to extend security specification to provide

- Trusted communications over 5G. The eavesdroppers are a big challenge
- Flexible and scalable security architecture
- Energy-efficient security – billions of IoT devices are unable to employ computational security solutions (lightweight solutions are required) [33].

Some of the important vertical services for the 5G includes Vehicle to Everything (V2X) and Internet of Things (IoT)

5G comes with architecture enhancements like

- Network Slicing (Logical Networks)
- Control Plane
- Service Based Architecture (SBA)
- Flexible non-3GPP access internetworking

5G are no longer monolithic network entities, a new authentication framework is needed to adapt to the change.

The current 3GPP standard focus on the next threats [34]

- Termination point of user plane (UP) security
- Authentication and authorization (Identity Management)
- RAN security
- Network slicing security
- Enhanced International Mobile Subscriber Identity (IMSI) privacy
- Increased home control

With massive usage of billions of IoT devices, some data will be just simple measurements like the temperature, while other data will be sensitive and critical. Applying the same security level on these different data is a clear waste of time and resources. So, applying multi-level security for 5G IoT devices will be a required feature for 5G IoT big data.

In [35], the authors propose a MLS (multi-level security) model, it's a state machine based model (properties & rules) that extends the BLP (Bell-LaPadula) model to be used in the information security domain.

5G is faster than 4G, more responsive, uses less power, gives strong and fast service more reliably and can connect more devices.

Main security concerns are

- Decentralized security (more traffic routing points)
- More bandwidths will be high pressure on the current security monitoring tools.
- Many IoT devices are manufactured with lack of security.
- Lack of early encryption in connection process (getting information about the device hardware & software by the attacker may help in planned attacks).

Common attacks are

- Botnet attacks
- DDoS
- MITM
- Location tracking and call interception

[36]

IoT devices are great target for cyberthreat, and this challenge increased by more use cases like connected cars and healthcare

In a survey, 94% of industry experts are worried that the security challenges will be increased by 5G networks. One of the security risks is related to switching between networks. Because the protocol designed to allow 4G or 3G connections when a 5G signal is not available [37]. The security requirements of 5G IoT devices encouraged companies in the industry to provide security solutions that match the needs of this technology [38]. 5G uses IMSI (International Mobile Subscriber Identity) encryption. All the traffic data sent over the Radio Network is encrypted [39].

Industry 4.0 is the large automation of factories using machine to machine communication and IoT devices [40-41]. The number of IoT connections is expected to be increased by 27% every year reaching 4.1 billion in 2024. Massive IoT cellular technologies includes [42]

* NB-IoT (Narrow band Internet of Things)

* Cat-M1 (LPWAN – Low Power Wide Area Network)

Mobile Network Operators (MNO) need to adopt comprehensive end-to-end security strategy [43].

- Complete visibility, inspection & control
- Cloud based threat analytics (Big data and Machine Learning)
- Security functions integrated with open application programming interface (API)
- Contextual security outcomes (To isolate infected devices)

## 7. 6G Networks

6G is the planned successor to 5G mobile networks. Many companies like Nokia, Samsung & LG and many countries like China & Japan are interested in 6G and many researchers are working on it. It is expected that by 2035, 6G will become able to send signals at the rate of human computation. 6G uses frequencies above 95 GHz, and it is expected that 6G phones could come with more advanced features, it may see behind the walls, replace eyeglasses (using goggles), provide very accurate position (down to cm) [44]. As an example about the huge interest in 6G, The Ministry of Science and Technology in China started two groups for the development of 6G, one by the government to determine how this will be carried out and the other group by 37 universities and enterprises [45]. It is expected that the 6G speed will be very huge, it could be 1TB per second, or 8 TB per second [46].

The key applications of 6G includes [47]

(1) Multi-sensory XR applications
(2) Connected robotics and autonomous systems
(3) Wireless brain-computer interaction
(4) Blockchain and distributed ledger technologies

The key technologies of a 6G includes [47]

(1) AI
(2) Molecular Communication
(3) Quantum Communication
(4) Blockchain
(5) THz (TeraHertz Technology)
(6) VLC (Visible Light Communication)

The key components of a 6G network [47]

(1) Real-time Intelligent Edge (real time response)
(2) Distributed AI (make intelligent decisions)
(3) Intelligent radio (self-adaptive) – Transceiver algorithms could dynamically configure and update themselves based on hardware information
(4) 3D intercoms (Full 3D Cover). coverage not only at the ground level but also at the space and the undersea level.

Edge Computing is about having the computation and data storage near to the location that need it (An idea like the idea behind the Content Delivery Network - CDN). An edge could be an IoT device, smartphone, self-driving car, etc. For example, in self-driving cars the vehicle cannot wait seconds for cloud processing because an accident may happen. It must process that data instantly and decide.

Toyota predicts that the car-to-cloud data stream will reach 10 exabytes per month by 2025 which is very huge [48].

6G networks will employ artificial intelligence (AI) to optimize and automate their operation. This could include the switch from mobile edge computing to AI at the Edge. i.e. Introducing intelligence at the edge devices.

AI at the Edge includes:

(1) Training data distributed over large number of edge devices.

(2) Each edge has fraction of the data.

(3) Reduction of network data (Massive amount of monitored data cannot be stored)

In distributed and federated AI, The IoT devices could share the same ML model and periodically exchange their neural network (NN) weights and gradients [49]. Applying Machine Learning in 6G mobile networks is a double-edged sword. We must be careful about Attacks on ML models and Violation via these models.

The attacker could

(1) Change the training data

(2) Get the training data & model

(3) Making adversarial examples (Misprediction)

(4) Inferring whether a sample belongs to a model

In Edge Computing we need to protect mobile device data privacy to prevent the attackers from misusing the local data [50].

The important 6G Characteristics are

* DL data rate > 1000 Gbps

* U-plane latency < 0.1 ms

* C-plane latency < 1ms

* Mobility up to 1000 km/h

* DL spectral efficiency - 100 bps/Hz

* Operating Frequency Up to 1000 GHz

From these characteristics, we will have a very huge traffic and AI could be used for modeling for malicious traffic detection in 6G networks. So, authentication can be based on Traffic & Requests analysis [51].

6G Key Performance Indices (KPIs) [52]

(1) Common KPI

    a. Capacity
    b. Spectrum Efficiency
    c. Energy Efficiency
    d. Intelligence

(2) Distinctive KPI

    a. Data Rate
    b. Latency
    c. Connectivity
    d. Security

6G is used for the interconnection of physical, biological, and digital worlds through the next operations

(1)     Precision sensing and actuation.
(2)     Ubiquitous compute and AI.
(3)     Multi-sensory rendering.
(4)     Human-Machine Interface.

The devices will radically evolve

(1) Create new digital worlds

    a. High resolution mapping.
    b. Mixed reality co-design.
    c. Holographic Telepresence.

(2) Augment our intelligence

    a. Learn from/with machines.
    b. Automatic security.
    c. In-body monitoring.

(3) Control the automatons

    a. Domestic robots.
    b. Remote & Self driving.
    c. Drone/robot swarms.

Spectrum options for 6G

1. Low Rank channel (30 GHz - 300 GHz) - mm

    * New 6G Sub Tera Hz Frequencies

    * Improved MIMO spectral Efficiency

2. High Rank Small Antennas (3 GHz - 30 GHz) - cm

* Improved Spectrum Utilization

3. Large Antenna Elements - (Less than 3 GHz) - dm

* AI based spectrum access

[53]

The number of connected devices is expected to be 45 billion by 2030 which mean an increase in the demand for stability, bandwidth, and latency. Also, Effective power allocation and distribution for 6G network in a box (6G-NIB) to enable peer to peer wireless communication networks is required. We can have optimized power allocation and distribution system using the Quasi-Convex Problem-Solving approach of Serial Polynomial Programming (SPP).

The SPP-based power optimization principle greatly enhances 6G -NIB and enhance Internet of Things (IoT) network efficiency and performance effectively.

[54]

Some researchers expect that 6G will enter the market by 2026. Key requirements by 6G includes

(1) optical free-space indoor communications,

(2) wireless charging and energy harvesting

(3) extensive use of machine learning to facilitate innovative services.

An interesting question is: What if future smartphones could be powered for a whole week/month without charging the battery?

This is expected in the 6G era [55].

In 6G Networks, the Satellite Communication at Millimeter Waves is a key enabler of 6G era.

Computing stations will be placed on the satellite platforms for complementing the terrestrial networks. We need high-capacity satellite communications to satisfy the requirements of huge traffic demands.

The satellites will provide

(1) service boosting for users in crowded areas

(2) eMBB (enhanced mobile broadband) in unserved and disaster areas

(3) Multi-connectivity for service continuity

The elevation angle is a key parameter in multibeam satellite communications because during the setup of satellite scenarios at mmWaves, using low elevation angles increase probability of path blockage and the more severe impact of atmospheric and aspects [56].

# 6G IoT Security

6G IoT still under research, but several names are called on the expected technology like Internet of Intelligence/Sense/Emotions to describe how it will look like. There are many problems that needs to be solved. That are related to Security, Privacy, Complexity & Sovereignty.

Major problem while enabling 6G with IoT is security because there is no standard exist and the most of users are not used to connecting IoT devices. [57]

"The 6G is about the 6th sense", as the network uses AI to perform actions in different situations as the 6G radio will sense the environment to connect the users [58]

In one of the studies:

(1) 90% of users do not trust the IoT due to Security

(2) 63% think that connected devices are creepy

[59]

Blockchain decentralized nature provide useful mechanism to deal with IoT challenges but Blockchain protocols with IoT failed to consider the computational loads, delays, and bandwidth overhead which lead to new set of problems.

The benefits for 6G includes Intelligent Resource Management but the Research Challenges exist:

(1) Huge number of IoT devices

(2) Energy Consumption

(3) Lack of Standardization

(4) Minimal Computational Power

(5) Low Storage Capacity

Benefits of blockchain with 6G-enabled IoT for industrial automation.

(1) Trust Building between 6G IoT devices and end users.

(2) Cost Reduction

(3) Fast Transaction

[60]

With respect to the IoT architecture and describing the different layers, we already have multiple opinions proposed by the researchers. A common one is the Perception, Network, and Application layers (According to many researchers)

- The Perception layer is related to the Sensors
- The Network layer is about data routing and transmission
- The Application layer is related to the applications and the authentication, integrity, and the confidentiality of the data

Attacks in the Network Layer could be

(1) DoS attacks

(2) Confidentiality & Privacy (Traffic analysis, Eavesdropping & Passive monitoring)

(3) Man-In-The-Middle attack

We have two types of communication to the internet

(1) Machine-to-Human

(2) Machine-to-Machine

We need to have security at the level of the Objects and the Network too [61].

Data Link Layer Communication Protocols in IoT [62]

1 - Bluetooth - (3 Mbps in a range of 50m to 150m)

2 - ZigBee - (10-100m)

3 - BLE (Bluetooth Low Energy)

4 - Wi-Fi (Wireless Fidelity) - data rate varies from 2Mbps to 1.73Gbps

5 - Z-Wave - 9.6Kbps, 40Kbps, or 100Kbps. (from 98 to 328 feet)

6 - RFID (Radio-Frequency Identification) - frequency bands of 13.56

7 - Cellular (GSM/3G/4G/5G etc).

8 - Sigfox - Large IoT networks (sets up antennas on towers)

9 - Ethernet - data transfer rates as high as 100 Mbps

10- NFC (Near Field Communication) - 13.56 MHz frequency.

11- LPWAN (Low Power Wide Area Network) - range varies from 2 km to 1000 km

12- LoRaWAN (Long Range Wide Area Network) - range of more than 15 km.

In [63], the authors propose IoT architecture based on 5 layers

(1) Physical sensing layer

(2) Network/Protocol layer

(3) Transport Layer

(4) Application Layer

(5) Data and Cloud Services

For the network layer they talked about

Attacks and the Security Countermeasures

    (1)    DDoS

* Ingress/Egress filtering, D-WARD, Hop Count Filtering and SYN-Cookies.

    (2) MITM

* Intrusion-detection system (IDS) and virtual private network (VPN).

    (3) Replay attacks

* Timeliness of Message.

And the solutions including usage of algorithms like

RSA, 3DES, AES, DSA & ECDH

They provided the algorithm used for each protocol, and the energy consumption. Also, they talked about the Lightweight encryption algorithms for IoT.

* Symmetric (PRESENT & CELFIA)

* Asymmetric (RSA, Elliptic Curves)

Energy Efficient Communication in 6G IoT is required. Also, the Customer satisfaction/perception (quality of experience (QoE)) is an essential factor to be analyzed.

The challenges include:

(1) High Speed and more Data Rates

(2) Heterogeneity and interoperability

(3) High energy and entropy drain

(4) Limited spectrum and high congestion The

[64]

IoT application programming tools include

(1) Node-RED (Developed using Node.js)

(2) ioBroker

(3) Flogo

(4) Kura (Eclipse - extensible open source IoT Edge Framework)

(5) Modbus

We have Network layer encapsulation protocols

(1) 6LoWPAN

(2) 6TiSCH

(3) ZigBee IP

(4) IPv6 over G.9959

(5) IPv6 over Bluetooth Low Energy

(6) IPv6 over NFC

(7) IPv6 over MS/TP-(6LoBAC)

(8) IPv6 over DECT/ULE

(9) IPv6 over 802.11ah

Network Layer Routing Protocols

(1) RPL (Distance Vector Routing Protocol for Low Power and Lossy Networks)

RPL Enhancements

* point-to-point reactive RPL (P2P-RPL)

* Enhanced-RPL

* Dynamic RPL

* mRPL

* Smarter-HOP (mRPL++)

(2) CORPL

* Every node keeps a forwarding set instead of its parent only.

(3) CARP (Channel-Aware Routing Protocol)

* Used in Underwater Wireless Sensor Networks (UWSNs)

* E-CARP allows the sink node to save previously received sensor data.

(4) AODV, LOADng and AODv2

* Ad Hoc On-Demand Distance Vector Routing

* Lightweight On-Demand Ad Hoc Distance Vector Routing Protocol-Next Generation (LOADng) and

* AODVv2. Contrary to AODV which just uses hop-count as a routing metric.

Routing protocol requirements includes:

(1) Match the traffic pattern of its deployment

(2) Resourceful in terms of power consumption

(3) Scale (Memory & Performance)

(4) Cope with sparse location changes

(5) Avoid one-way links

 6) Support IPv6

[65]

Cognitive Routing uses Machine Learning (ML) algorithms to optimize routing decisions. Most of the research is about traffic prediction and route optimization. In [66] The authors propose a cognitive routing framework for KDN (Knowledge-Defined Networks). The framework uses a Shortest Path Algorithm names MRoute, that proactively computes all-possible paths between all pairs of nodes. Further, it uses Sharpe-Ratio to measure volatility of each link and RNN (Recurrent Neural Network) with LSTM (Long Short-Term Memory) to learn trend. The framework uses online learning to tackle any dynamic network behavior.

## 8. Programming Without Coding Technology

Developing IoT software requires different programming skills and different programming languages. This causes a problem for companies that need to hire many programmers with different skills. The PWCT visual programming tool increase productivity and provide researchers with one visual programming tool to develop different solutions. PWCT support code generation in many programming languages like C, Python, and Harbour [67-69]. PWCT support the Super Server programming paradigm [70]. It's used in the first implementation of the LASCNN algorithm and in developing many business applications [71-72]. Also, it's used for creating the Supernova programming language and the Ring programming language [73-74]. Ring is a dynamic programming language like Python and Ruby, comes with small implementation like Lua and an IDE like Visual Basic. The language comes with innovative features related to building domain-specific languages using declarative programming and natural language programming. It's used in many GUI and front-end applications [75-76]. As in Social Networks [77], security in IoT is very important too, and with billions of devices everywhere, it's more important and more critical.

## 9. Conclusion

The last few years have witnessed a huge growth in the wireless telecommunications. In this paper we review the mobile network generations from the first generation to the six generation and we present the security challenges related to IoT applications. We highlighted the security challenges that are more threating in 6G IoT unless properly addressed.

A general and standard security framework is required to provide a trusted solution for these challenges. This framework could use different ideas and technologies like artificial intelligence, block-chain, multi-level security systems and standard development tools, algorithms and libraries that are designed with security in mind.